**Guises and Perspectives: An Intentional and Hyperintensional Sketch**

Author: Juan J. Colomina-Alminana

Affiliation: Northeastern University, Oakland Campus (Mills College)

Address: 5000 MacArthur Blvd., 132 Vera Long, Oakland, CA, 94613 – USA

Email: j.colomina-alminana@northeastern.edu

Declaration: The author declares that has no conflict of interest.

No data employed

This article adheres to all ethical standards

**Abstract:** This paper develops a formal logic for guises based on Héctor-Neri Castañeda's account, who understood relations from an internalist viewpoint, following Leibniz's work. We introduce a syntax, model theory, and proof theory for an intensional logic in which guises—bundles of properties equipped with intention—serve as primary semantic objects. The system integrates (i) a Leibnizian containment semantics for singular truths, (ii) an intentional operator that captures internal relations among guises, and (iii) a modal layer for possibility and necessity modeled as maximally consistent closures. We establish core metatheoretic results (e.i. soundness and canonical-model completeness sketches) and analyze hyperintensional phenomena such as substitution failure in intentional contexts, quasi-indexicality, and *de se* reference. We compare the framework to classical intensional semantics (Montague), property theory (Bealer), hyperintensional logics (Fine), situation semantics (Barwise and Perry), and to the Leibniz's program for a calculus of concepts. The result is a self-contained formal framework that demonstrates that relations are not external causal links but intentional internal structures encoded in the guises through which agents and objects are conceived—i.e., they are perspectives.

* * *

Castañeda's (1975a) guise theory treats individuals, fictional objects, and even the self as guises. These are interpreted as property bundles that can be presented under different conceptual profiles, and that can intentionally refer to or aim at other guises. The theory was motivated by Castañeda's account to indexicals, quasi-indexicals (terms such as "he himself"), *de se* attitudes, and the logic of self-ascription, where purely extensional substitution principles fail.[1] In parallel, Leibniz famously grounded singular truths in predicate containment in a subject's complete concept and explained inter-substantial coordination via internal representation rather than causal connections. In this article, we develop a system that synthesizes these lines: Guises are subjects of containment, intention is the internal mapping that yields a logic of relations, modal possibility is consistency, and necessity is inclusion across maximal closures. These are, we argue, the parameters that forge perspectives.

The aim of this paper is twofold. First, we articulate a formal logic of guises that captures Castañeda's program. It builds a model for intentional reference, *de se* thought, and cross-perspective identity not by positing exotic objects, but by providing a rigorous semantics for bundles of properties and their internal relations. Second, we demonstrate how such logic completes a (Leibnizian) picture of relations as internal. Under this viewpoint, relational truths emerge from conceptual containment and intentional structure, not from external ties or causal transmission. The overarching claim is methodological though. A logic of guise can be both metaphysically charitable to Leibniz and technically adequate for contemporary logic, preserving intensional distinctions where they matter while anchoring truth in extensional containment where it is safe to do so.

[1] Castañeda (1966: 145) says: "There are contexts in which the substitution of coreferential singular terms fails to preserve truth."

We proceed as follows. Section 1 situates the framework. Section 2 gives the language and core operators. Section 3 provides the model theory: Guises, properties (marks), closure (complete concepts), intention sets, and worlds. Section 4 presents axioms and rules, including intention–closure alignment principles and a relational operator R for internal relatedness. Section 5 sketches soundness and canonical-model completeness. Section 6 treats hyperintensionality, quasi-indexicals, and *de se* logics. Sections 7–9 develop extensions (epistemic, agency, dynamics), algebraic/categorical perspectives, and works case studies. Section 10 discusses expressiveness, complexity, and decidability. Section 11 addresses objections and failure cases. Section 12 offers some concluding remarks.

## 1. Historical background and motivations

Leibniz held that singular truths are grounded in predicate containment within a subject's complete concept. According to him, relations are "internal denominations," ideal features of concepts rather than external linkages among things.[2] As it is well known, Leibniz claimed that Monads have no windows, yet they represent the entire universe from their point of view.[3] This internalism invites a logic that derives relational structure from within conceptual profiles.

[2] See, especially, Leibniz's (1686) on containment doctrine and "internal denominations," in Garber and Ariew (1989).
[3] "The monads have no windows through which anything could enter or leave" (Leibniz 1714: §7). "Each created monad represents the whole universe according to its point of view" (Leibniz 1714: §56).

Castañeda, addressing puzzles of indexicals and attitude ascriptions, argued that we must distinguish conceptual guises under which objects are presented, otherwise extensional substitution fails systematically in *de se* and *de re* contexts.[4] He introduced quasi-indexicals to capture the phenomenon that one can believe of oneself, under a third-person guise, what one cannot yet endorse as I-thoughts. The lesson from this is, primarily, that internal presentation (guise) is constitutive of reference in intensional contexts.

Taking this as a base, we develop a system that carries forward three guiding ideas. First, predicate containment works as a truth-maker for singular predications, which as shown above can be traced back to at least Leibniz. Second, intentional operators are understood as governing internal reference among guises, as Castañeda directs. Third, we also include a modal layer where possibility is (theory-)consistency and necessity is inclusion across maximal closures, what is usually called Leibnizian ratiocination. In these terms, relations become constructions out of intention and containment, aligning with Castañeda's internalist metaphysics and Leibniz's exegesis. We are also in agreement with Quine when claiming that quantification fixes ontological commitment,[5] and with Montague in that natural language can be rendered in formal semantics.[6] We expand this commitment though to add a hyperintensional, guise-sensitive layer where substitution and identity must be constrained to preserve the phenomena Castañeda uncovered.

## 2. The Language for a Logic of Guise

We propose GL (Guise Logic), a many-sorted intensional language.

[4] Cf. Castañeda (1967). See especially the section on failure of substitution in *de se* contexts.
[5] "To be is to be the value of a bound variable" (Quine 1948: 22).
[6] "There is no important theoretical difference between natural languages and the artificial languages of logicians" (Montague 1970: 188).

I. Sorts: marks/properties (lowercase Latin p, q, r), guises (g, h, u, v), propositions (φ, ψ, χ) treated as sets of marks, and worlds (w).

II. Non-logical vocabulary:

- Unary predicates on marks may be omitted. We treat marks as primitive.
- A satisfaction symbol ⊨ for semantic metalanguage.
- A binary intentional operator $\mathcal{I}$(g, φ): "guise g intends proposition φ."
- A binary internal relation R(g, h): "g is internally related to h."
- A reference operator ρ(g, h): "g intentionally refers to h" (definable via $\mathcal{I}$ and satisfaction).
- Modal operators □, ◇ for necessity and possibility.

III. Connectives and quantifiers: ∧, ∨, →, ¬, ∀, ∃ over marks, guises, and propositions.

Schematic definitions within the object language will be supported by the model theory. For proof theory, we employ either a Hilbert system with typed quantification or a Gentzen sequent system. For readability we often write φ ⊆ g to mean "proposition φ (set of marks) is contained in guise g."

**3. The Semantics for GL**

We interpret GL on structures M = ⟨$\mathcal{P}$, G, W, T, κ, $\mathcal{I}$⟩, where:

I. $\mathcal{P}$ is a non-empty set of marks (properties, "notes" in Leibniz's sense).

II. G ⊆ $\wp(\mathcal{P})$ is a non-empty set of guises, each a set of marks (bundles).

III. T is a background theory over $\mathcal{P}$ (a set of conditionals among marks) inducing a Tarskian consequence operator Cn_T with extensivity, monotonicity, and idempotence.

IV. κ: $\wp(\mathcal{P}) \to \wp(\mathcal{P})$ is the closure operator κ(X) = Cn_T(X), modeling "complete concepts."

V. W is a non-empty set of worlds, each $w \subseteq \mathcal{P}$ that is T-closed and (typically) maximal consistent.

VI. $\mathcal{I}: G \rightarrow \wp(\wp(\mathcal{P}))$ assigns to each guise g a family $\mathcal{I}(g)$ of intended propositions (sets of marks).

Satisfaction is defined as:

I. $g \vDash \varphi$ iff $\varphi \subseteq g$ (containment).

II. $w \vDash \varphi$ iff $\varphi \subseteq w$ (world containment).

Internal relation and reference:

I. $R(g, h)$ iff $\exists\varphi \in \mathcal{I}(g)$ with $\varphi \subseteq h$.

II. $\rho(g, h)$ is definable as $R(g, h)$ with the additional constraint that $\varphi$ determines h up to an equivalence (see §6 below for hyperintensional refinements).

Modal truth:

I. $M \vDash \Diamond\varphi$ iff $\exists w \in W$ with $w \vDash \varphi$.

II. $M \vDash \Box\varphi$ iff $\forall w \in W, w \vDash \varphi$.

Leibnizian truth for singular predication P(g) is rendered as $P \in \kappa(g)$: The predicate/mark P is contained in the complete concept of g. This captures containment as truthmaker.

Two guiding design choices make the semantics work:

I. Intention–closure alignment: $\mathcal{I}(g)$ should be stable under T-consequence (if $\varphi \in \mathcal{I}(g)$ and $T \vdash \varphi \Rightarrow \psi$, then $\psi \in \mathcal{I}(g)$).

II. Extension invariance: if $g \subseteq h$ (or $\kappa(g) \subseteq \kappa(h)$), then $\mathcal{I}(g) \subseteq \mathcal{I}(h)$. This reflects that completion or enrichment of a guise should not destroy its intentions.

We do not force these as semantic necessities. Instead, we axiomatize them and study failure cases (see §11 below).

## 4. Axioms and rules

We present a Hilbert-style system $GL^+$ with typed quantifiers and a primitive binary predicate Int(g, φ) for $\mathcal{I}(g, \varphi)$. Derived forms use set-theoretic notation in the metalanguage.

Core axiom schemata:

I. Containment Truth (CT): If $\varphi \subseteq g$, then $g \vDash \varphi$.

II. Leibniz Containment (LC): For any mark P and guise g, $\Box(P(g) \leftrightarrow P \in \kappa(g))$.

III. Intentional Connection (IC): $R(g, h) \leftrightarrow \exists\varphi\ (\mathrm{Int}(g, \varphi) \land \varphi \subseteq h)$.

Intention–closure principles:[7]

I. CI1 (Consequence Closure): $\mathrm{Int}(g, \varphi) \land (\varphi \Rightarrow_T\ \psi) \rightarrow \mathrm{Int}(g, \psi)$.

II. CI2 (Extension Invariance): $g \subseteq h \rightarrow (\mathrm{Int}(g, \varphi) \rightarrow \mathrm{Int}(h, \varphi))$.

III. CI3 (Commutation): $\kappa\char`^\diamond(\mathcal{I}(g)) = \mathcal{I}(\kappa(g))$, where $\kappa\char`^\diamond$ lifts closure to families of propositions: $\kappa\char`^\diamond(S) = \{\psi : \exists\varphi_1,\ldots,\varphi_n \in S \text{ with } T \vdash \bigcup_i\ \varphi_i \Rightarrow \psi\}$.

Relational axioms:

I. R1 (Reflexivity under Self-Intention): $\mathrm{Int}(g, \varphi) \land \varphi \subseteq g \rightarrow R(g, g)$.

II. R2 (Monotonicity): $h \subseteq h' \land R(g, h) \rightarrow R(g, h')$.

III. R3 (Conditional Transitivity): $(\mathcal{I}(h) \subseteq \mathcal{I}(g)) \land R(g, h) \land R(h, u) \rightarrow R(g, u)$.

Modal axioms (Leibnizian):

I. M1 (T-Closed Worlds): If $\varphi \subseteq w$ and $T \vdash \varphi \Rightarrow \psi$, then $\psi \subseteq w$, for all $w \in W$.

II. M2 (Possibility = Consistency): $\Diamond\varphi \leftrightarrow \exists w\ (w \vDash \varphi)$.

III. M3 (Necessity = Universal Inclusion): $\Box\varphi \leftrightarrow \forall w\ (w \vDash \varphi)$.

IV. From non-emptiness of W: $\Box\varphi \rightarrow \Diamond\varphi$.

---

[7] Cf. Bealer (1982), Fine (2012), Fine (2017), and Restall (2000).

Rules:

I. Modus Ponens, Generalization over appropriate sorts.

II. From Int(g, φ) and φ ⇒_T ψ infer Int(g, ψ) (CI1-rule).

III. From g ⊆ h and Int(g, φ) infer Int(h, φ) (CI2-rule).

IV. From Int(g, φ) and φ ⊆ h infer R(g, h) (IC-rule).

We assume standard propositional and first-order axiomatics. The identity theory can be added with Leibnizian identity by closure (κ(g) = κ(h) → g ≡ h extensional identity) plus a guarded substitutivity for non-intentional contexts.

**5. Soundness and Completeness**

Soundness follows by inspection from the model clauses.

I. CT and LC are immediate from containment and definition of κ.

II. IC is by definition of R.

III. CI1 holds if $\mathcal{I}$(g) is closed under T-consequence. We either enforce it semantically or treat it as an axiom constraining admissible models.

IV. CI2 holds under extension invariance of intention (either semantic or axiomatic).

V. CI3 holds in canonical models where $\mathcal{I}$(g) = {φ : φ ⊆ κ(g)} or, more generally, where $\mathcal{I}$ depends only on κ(g) and is T-closed.

VI. R1–R3 follow from IC plus set inclusion properties.

VII. M1–M3 are standard if worlds are maximal T-closed sets.

Canonical-model completeness. Define the Lindenbaum algebra of $GL^+$-theories. From there, one can build a canonical model whose guises are equivalence classes of guise-terms *modulo* provable equivalence, with $\mathcal{I}$(g) set to the deductively closed set of φ such that Int(g, φ) is in the theory.

Worlds are maximal consistent, T-closed sets of marks. Truth lemmas are proved by induction on formulas. It is important to notice that, here, the key steps require CI1–CI3 to ensure that the canonical $\mathcal{I}$ interprets axioms correctly. Completeness, then, follows for the intended class of models (those satisfying CI1–CI3 and M1–M3).

We leave a full Henkin construction and filtration arguments to an extended version. The finite model property can be obtained for fragments (i.e., Horn-T, finite Θ-templates. See §9 below), yielding decidability for those fragments.

## 6. Hyperintensionality, Quasi-Indexicals, and *de se* Attitudes

Castañeda's central data concern intensional contexts where coreference does not license substitution. In $GL^+$, substitutivity fails in Int(–, –) and R(–, –) positions by design. Even if $\kappa(g) = \kappa(h)$, it can be that $\mathcal{I}(g) \neq \mathcal{I}(h)$, so Int(g, φ) does not need to imply Int(h, φ). This models cognitive significance and *de se* asymmetries. We make this precise.

I. Definition (Hyperintensional non-collapse). A model M is hyperintensional if $\exists g, h$ with $\kappa(g) = \kappa(h)$ and $\mathcal{I}(g) \neq \mathcal{I}(h)$. In such models, for some context C[·], C(g) holds but C(h) fails. This is exactly what we can expect in quasi-indexical constructions, as predicted by Castañeda (1967). Cf. Castañeda (1966) and Prior (1971).

II. Quasi-indexicals. To capture "he himself," we can enrich the language with a *de se* operator Self(g, φ), read "g self-ascribes φ." Semantically, Self(g, φ) holds iff Int(g, φ) and $\varphi \sqsubseteq g$. This recovers R(g, g) and provides a neat formalization of self-ascriptive knowledge.[8]

[8] Castañeda (1967: 91) says: "The individual identifies himself under a guise; the guise figures in the content of what is thought." Cf. Barwise & Perry (1983), Duží *et al.* (2010), and Yablo (2014).

III. *De re/de dicto*. Read Int(g, φ) *de dicto*. Add a *de re* operator Int^⋆(g, h, P) for "g intends of h that P," defined as $\exists\varphi \in \mathcal{I}(g)$ with $\varphi \subseteq \kappa(h)$ and $P \in \varphi$. This bridges *de re* attitudes to Leibnizian containment.[9]

Bonus: Hyperintensional upgrades. We can refine $\mathcal{I}$ to track proof-theoretic strength (Fine-style) or truth-makers: Let $\mathcal{I}(g)$ be closed under strict consequence but not under logical equivalence, to preserve differences between coextensive but non-identical propositions. This matches the phenomenology of belief ascription and complements the containment backbone.

**7. Extensions: Epistemic, Agency, Dynamics**

I. Epistemic operators K_g(φ). Add K_g capturing knowledge of g. Semantics: K_g(φ) iff for all epistemic alternatives $e \in E(g)$, $e \vDash \varphi$ and $\varphi \in \mathcal{I}(g)$. This ties knowledge to intention (endorsed contents) and to alternative sets. *De se* knowledge becomes K_g(Self(g, φ)).

II. Agency (STIT-style). Introduce [g stit: φ] for "g sees to it that φ."[10] We can constrain agency by intentions: [g stit: φ] → Int(g, φ). Combine with Leibnizian closure to obtain a calculus of intentional action where success conditions are containment in worlds resulting from choice functions.

III. Dynamics (public announcements). An update operator [!φ] transforms both W and $\mathcal{I}$: Worlds are restricted to those satisfying φ, while intentions update by CI1 reflect consequences of φ. This yields a dynamic guise logic that models conceptual change and narrative identity.

---

[9] Cf. Soames (2002), where the key difference with the model here presented is the mandate for structured propositions.
[10] Cf. Belnap *et al.* (2001), which provides the semantics and proof theory for agentive modalities in indeterministic settings.

IV. Situation semantics. Replace worlds w with situations s (partial $\mathcal{P}$-sets), interpret Int against situations, and let containment be relative to s. We reclaim Barwise–Perry expressivity while retaining the guise machinery.

## 8. Algebraic and Categorical Semantics

I. Concept lattice. The family of T-closed sets of marks ordered by ⊆ forms a complete lattice. Guises embed into this lattice via κ. $\mathcal{I}(g)$ can be modeled as a principal ideal $\downarrow\kappa(g)$ or as a Θ-filter (see §9 below) to keep intentions finite. The internal relation R is then a Galois connection: Define $F(g) = \mathcal{I}(g)$, $U(h) = \{\varphi : \varphi \subseteq h\}$. Then $R(g, h) \leftrightarrow F(g) \cap U(h) \neq \emptyset$. Algebraic properties (i.e. monotonicity, residuation) fall out.

II. Categorical gloss. Let C be a category whose objects are T-closed concepts and whose morphisms are inclusions. An "intention functor" I: C → Pos (posets) maps a guise g to its poset of intended propositions. Natural transformations from I to the Yoneda embedding y: C → Set pick out canonical intention sets (downsets). This categorifies CI3 as a naturality condition.

## 9. Worked Models and Templates

I. System A (canonical): $\mathcal{I}(g) = \{\varphi : \emptyset \neq \varphi \subseteq \kappa(g)\}$. This makes CI1–CI3 valid by construction. It is maximally intensional in that every included proposition is intended. It is too strong for cognitive modeling but ideal as a completeness anchor.

II. System B (template-restricted): Fix a finite, T-closed $\Theta \subseteq \wp(\mathcal{P})$ of role templates. Define $\mathcal{I}(g) = \{\theta \in \Theta : \theta \subseteq \kappa(g)\}$. This keeps intentions structured and finite, often yielding decidable fragments (Horn T, finite Θ). R(g, h) reduces to existence of a role θ jointly satisfied by g and h.

III. System C (finite example): $\mathcal{P}$ = {a, b, c, d}, T has a ⇒ b, b ∧ c ⇒ d. Compute closures and intentions explicitly, and verify axioms and derive sample theorems (monotonicity, necessity ⇒ possibility, conditional transitivity) as in standard algebraic semantics. This serves as a laboratory for checking counterexamples (see §11 below).

**10. Expressiveness, Decidability, Complexity**

I. Fragments. The purely propositional fragment with Horn T and finite Θ is decidable (PSPACE at worst. Often, PTIME via forward chaining on T and membership checks on Θ). Adding first-order quantification over marks and guises raises complexity, whereas guarded fragments remain decidable. Adding STIT or dynamic operators typically preserves decidability under finite models.

II. Finite model property (FMP). For Horn T and finite Θ, filtration yields FMP. Hyperintensional refinements without closure under equivalence preserve filtration provided syntactic shape is constrained.

III. Comparative expressiveness. $GL^+$ strictly extends standard Kripke–Montague intensional logics by admitting hyperintensional intention contexts and *de se* operators, while retaining a Leibnizian extensional backbone for singular predications (LC). It is comparable to Bealer's (1982) property theory in modeling properties directly but differs by (i) the guise-as-bundle base type, (ii) the intention operator as primitive, and (iii) the built-in closure κ for complete concepts. In addition, it is complementary to Fine's (2017) truth-maker semantics: $\mathcal{I}$ can be interpreted as a set of admissible truth-makers for g's contents.

## 11. Objections, Failure cases, and Repairs

We summarize instructive failure modes and accompanying axioms.

I. Non-commutation (κ versus $\mathcal{I}$). If $\mathcal{I}$ depends on non-conceptual features (i.e. memory, attention), CI3 may fail: $\kappa^\diamond(\mathcal{I}(g)) \neq \mathcal{I}(\kappa(g))$. Repair: Adopt CI3 where modeling a Leibnizian ideal agent, drop it in cognitive variants.

II. Non-transitive R. Without intention inheritance ($\mathcal{I}(h) \subseteq \mathcal{I}(g)$), there is no need for R to be transitive. Repair: Assume R3 only under inheritance. Otherwise, accept non-transitivity as capturing perspective shifts.

III. Exact-intention brittleness. If $\mathcal{I}(g)$ excludes logical consequences (hyperintensional), modal monotonicity ($\Diamond Int(g, \varphi) \rightarrow \Diamond Int(g, \psi)$ when $\varphi \Rightarrow_T \psi$) can fail. Repair: Add a weak consequence-closure (CI1′) for a selected base of consequences.

IV. Substitution failures. Even when $\kappa(g) = \kappa(h)$, $Int(g, \varphi)$ may differ from $Int(h, \varphi)$. We embrace this as a feature (hyperintensionality). In extensional contexts, we retain substitutivity via LC and standard identity axioms.

V. Identity. If two guises share κ but diverge in $\mathcal{I}$, are they "the same"? Can be substituted *salva veritate*? We distinguish extensional identity (κ-equality) from intensional profile identity ($\mathcal{I}$-equality). Both equivalences have logical roles. Therefore, neither collapses into the other.

These pressure points are, we believe, the phenomena Castañeda wanted logic to honor.

## 12. Conclusion

We have presented $GL^+$, a logic of guises that renders Castañeda's insights into a rigorous framework and aligns them with Leibniz's theory of (internal) relations, what we have argued was his interpretation and original intentions (Cf. also Castañeda 1975b, 1976, and 1990). Guises are bundles of properties. Relationality is internal, realized via intention. Truth in singular predication is predicate containment and modality is consistency and inclusion. The system is flexible: It admits canonical Leibnizian models (System A), cognitively realistic, finite role models (System B), and concrete finite laboratories (System C). It sustains hyperintensional distinctions essential for quasi-indexicals and *de se* thought whereas preserving an extensional backbone that makes derivations tractable.

Castañeda's guiding thought (the conceptual profiles, this is, the perspectives, under which we intend, refer, and act encode who and what we are) finds a precise home here, as demonstrated. Leibniz's dream of a complete calculus of concepts also finds here a renewed life: The logic of guises is nothing else that a calculus where inclusion, intention, and necessity fit together as parts of a single intentional architecture. After all, as Leibniz (1714: §9) said, "each individual substance expresses the whole universe in its own manner… [like] a living mirror of the universe."

**Appendix. Fully Developed Systems A–C (§10)**

Below are complete developments of Systems A–C from the §10 of our guise logic framework. Each system specifies the underlying ontology, closure theory, intention assignment, internal relation, modal layer, meta-results, and example derivations. They are designed to be self-contained and checkable, like any good finitist program requires, with increasing structure and constraints from A to C.

**System A: Canonical Closure–Intention System**

Ontology and theory

Marks: A non-empty set $\mathcal{P}$.

Guises: $G \subseteq \wp(\mathcal{P})$, each $g \in G$ a set of marks.

Theory: $T \subseteq$ {finite Horn implications over $\mathcal{P}$}. Define $\kappa(X) = Cn_T(X)$ (Tarskian closure: extensivity, monotonicity, idempotence).

Worlds: $W = \{w \subseteq \mathcal{P} : w$ is T-closed and maximal T-consistent$\}$.

Intention assignment

Definition: $\mathcal{I}(g) = \{\varphi \subseteq \kappa(g) : \varphi \neq \emptyset\}$. Intention is the non-empty downset of $\kappa(g)$.

Immediate properties:

CI1 (consequence closure): If $T \vdash \varphi \Rightarrow \psi$ and $\varphi \in \mathcal{I}(g)$, then $\psi \subseteq \kappa(g)$ hence $\psi \in \mathcal{I}(g)$.

CI2 (extension invariance): $g \subseteq h \Rightarrow \kappa(g) \subseteq \kappa(h) \Rightarrow \mathcal{I}(g) \subseteq \mathcal{I}(h)$.

CI3 (commutation): $\kappa\text{^}\diamond(\mathcal{I}(g)) = \mathcal{I}(g)$ and $\mathcal{I}(\kappa(g)) = \mathcal{I}(g)$.

Internal relation and reference

Definition: $R(g, h) \leftrightarrow \exists\varphi \in \mathcal{I}(g)$ with $\varphi \subseteq h$.

Witness criterion: Equivalently, $R(g, h)$ iff $h \cap \kappa(g) \neq \emptyset$ (pick $\varphi = \{p\} \subseteq h \cap \kappa(g)$).

Modal layer

Truth at worlds: $w \vDash \varphi$ iff $\varphi \subseteq w$; $\Box\varphi$ iff $\forall w \in W, w \vDash \varphi$; $\Diamond\varphi$ iff $\exists w \in W, w \vDash \varphi$.

Necessity implies possibility: If $\Box\varphi$ and $W \neq \emptyset$ then $\Diamond\varphi$.

Metatheory

Soundness: Trivial from definitions.

Completeness (sketch): Canonical model with $\mathcal{I}(g)$ as the deductive downset of $\kappa(g)$ yields truth lemmas. The pure propositional fragment is compact and has the finite model property if T is finite Horn.

Substitution: Extensional contexts permit κ-equality substitution. Intensional contexts (Int, R) collapse to extensional overlap under this system. Therefore, fewer hyperintensional distinctions (by design).

Worked example

Let $\mathcal{P} = \{a, b, c\}$; $T = \{a \to b\}$. Then:

$\kappa(\{a\}) = \{a, b\}$; $\kappa(\{b\}) = \{b\}$; $\kappa(\{c\}) = \{c\}$.

$\mathcal{I}(\{a\})$ = all non-empty subsets of $\{a, b\}$.

$R(\{a\}, \{b, c\})$ holds (witness $\{b\}$); $R(\{b\}, \{a\})$ fails.

**System B: Template-Restricted Intention System**

Ontology and theory

As in System A, but fix a finite template base $\Theta \subseteq \wp(\mathcal{P})$ that is T-closed:

If $\theta \in \Theta$ and $T \vdash \theta \Rightarrow \psi$ and $\psi \subseteq \mathcal{P}$, then $\psi \in \Theta$ (closure within $\Theta$).

Intention assignment

Definition: $\mathcal{I}(g) = \{\theta \in \Theta : \theta \subseteq \kappa(g)\}$.

Properties:

CI1: By Θ's T-closure, $\theta \in \mathcal{I}(g)$ and $\theta \Rightarrow_T\ \psi$ with $\psi \in \Theta$ implies $\psi \in \mathcal{I}(g)$.

CI2: $g \subseteq h \Rightarrow \kappa(g) \subseteq \kappa(h) \Rightarrow \mathcal{I}(g) \subseteq \mathcal{I}(h)$.

CI3: $\kappa\hat{}\diamond(\mathcal{I}(g)) = \mathcal{I}(g)$ because every consequence in Θ already belongs to $\mathcal{I}(g)$. Also, $\mathcal{I}(\kappa(g)) = \mathcal{I}(g)$.

Internal relation and reference

Definition: $R(g, h) \leftrightarrow \exists\theta \in \Theta\ (\theta \subseteq \kappa(g) \wedge \theta \subseteq h)$.

Reading: g relates to h when they share a template θ sanctioned by T.

Modal and proof-theoretic advantages

Finite basis: Reasoning about Int and R reduces to membership and subset checks in finite Θ.

Decidability: With Horn T and finite Θ, satisfiability is decidable (polynomial-time closure computation plus Θ-membership).

Worked example

Let $\mathcal{P} = \{a, b, c, d\}$; $T = \{a \rightarrow b, b \wedge c \rightarrow d\}$; $\Theta = \{\{a\}, \{b\}, \{c\}, \{d\}, \{b, c\}\}$.

For $g_1 = \{a\}$: $\kappa(g_1) = \{a, b\}$; $\mathcal{I}(g_1) = \{\{a\}, \{b\}\}$.

For $g_2 = \{b, c\}$: $\kappa(g_2) = \{b, c, d\}$; $\mathcal{I}(g_2) = \{\{b\}, \{c\}, \{d\}, \{b, c\}\}$.

$R(g_1, g_2)$ holds (witness $\{b\}$); $R(g_2, g_1)$ holds (witness $\{b\}$).

**System C: Finite Explicit Model with Calculations**

Ontology and theory

Marks: $\mathcal{P} = \{a, b, c, d\}$.

Theory T: $a \rightarrow b$; $b \wedge c \rightarrow d$. Compute κ via forward chaining.

Worlds (T-closed): $w_1 = \{b\}$, $w_2 = \{a, b\}$, $w_3 = \{c\}$, $w_4 = \{b, c, d\}$, $w_5 = \{a, b, c, d\}$. Maximal T-consistent among these are $w_2$, $w_4$, $w_5$.

Intention options

We present two variants to illustrate trade-offs.

Variant C1 (canonical downset):

$$\mathcal{I}(g) = \{\varphi : \emptyset \neq \varphi \subseteq \kappa(g)\}.$$

Hyperintensionality: low (akin to System A).

Variant C2 (role templates):

$\Theta = \{\{a\}, \{b\}, \{c\}, \{d\}, \{b, c\}\}$ as in System B.

$\mathcal{I}(g) = \{\theta \in \Theta : \theta \subseteq \kappa(g)\}$.

Hyperintensionality: moderate, it distinguishes intentions that are not in $\Theta$.

We will use Variant C2 for concreteness.

Closure computations

$\kappa(\{a\}) = \{a, b\}$.

$\kappa(\{b\}) = \{b\}$.

$\kappa(\{c\}) = \{c\}$.

$\kappa(\{b, c\}) = \{b, c, d\}$.

$\kappa(\{a, c\}) = \{a, b, c, d\}$.

$\kappa(\{d\}) = \{d\}$ (no rule introduces d alone).

Intention sets (Variant C2)

$\mathcal{I}(\{a\}) = \{\{a\}, \{b\}\}$.

$\mathcal{I}(\{b\}) = \{\{b\}\}$.

$\mathcal{I}(\{c\}) = \{\{c\}\}$.

$\mathcal{I}(\{b, c\}) = \{\{b\}, \{c\}, \{d\}, \{b, c\}\}$.

$\mathcal{I}(\{a, c\}) = \{\{a\}, \{b\}, \{c\}, \{d\}, \{b, c\}\}$.

Internal relation instances

$R(\{a\}, \{b, c\})$ via $\theta = \{b\}$.

$R(\{b, c\}, \{a, c\})$ via $\theta = \{b, c\}$.

$R(\{c\}, \{b, c\})$ via $\theta = \{c\}$.

R({b}, {a}) via θ = {b}[11]

Modal truths

□{b} is false ($w_3$ = {c} lacks b).

◇{d} is true ($w_4$, $w_5$).

From □φ → ◇φ holds since W ≠ ∅.

Theorems verified

CI1: From {b, c} ⇒_T {d} and {b, c} ∈ 𝑱({b, c}) infer {d} ∈ 𝑱({b, c}).

CI2: {a} ⊆ {a, c} implies 𝑱({a}) ⊆ 𝑱({a, c}).

CI3: κ^⋄(𝑱({a, c})) = 𝑱({a, c}) (Θ is T-closed).

R1: If θ ∈ 𝑱(g) and θ ⊆ g then R(g, g). Example: θ = {b, c} ∈ 𝑱({b, c}) and {b, c} ⊆ {b, c}.

R2: If h ⊆ h′ and R(g, h) then R(g, h′). Example: R({a}, {b}) via {b}, since {b} ⊆ {a, b}, also R({a}, {a, b}).

R3: Suppose 𝑱(h) ⊆ 𝑱(g). Let g = {a, c} and h = {b, c}. Then, 𝑱(h) ⊆ 𝑱(g). If R(g, h) via {b, c} and R(h, u) via {c} with u = {c}, then R(g, u) via {c}.

Hyperintensional tuning

Add a second template set Θ′ that distinguishes {a} and "a-because-a" as separate templates (procedural forms). Then, 𝑱 becomes sensitive to proof paths, enabling substitution failure among coextensive intentions while preserving CI1 for a curated base of consequences.

[11] Notice that {b} ⊆ κ({b}) and {b} ⊆ {a} is not possible. Therefore, R({b}, {a}) fails. Nevertheless, realize that R({a}, {b}) holds via {b} ⊆ {b}.